\documentclass[useAMS]{mn2e}
\usepackage{graphicx}
\usepackage{color}
\usepackage{colortbl}

\title[Carbon, Nitrogen and Oxygen Abundances]
{Carbon, Nitrogen and Oxygen Abundances in Atmospheres of the 5-11 M$_\odot$
 B-type Main Sequence Stars}

\author[L.S. Lyubimkov et al.]
{Leonid~S.~Lyubimkov,$^1$\thanks{E-mail: lyub@crao.crimea.ua (LSL); dll@astro.as.utexas.edu (DLL)} David~L.~Lambert,$^2$$^\star$ Dmitry~B.~Poklad,$^1$ \and Tamara~M.~Rachkovskaya,$^1$ Sergey~I.~Rostopchin$^2$\\
$^1$Crimean Astrophysical Observatory, Nauchny, Crimea, 98409, Ukraine\\
$^2$The W.J. McDonald Observatory, The University of Texas at Austin, Austin, TX 78712-0259, USA\\}

\begin{document}

\date{Accepted. Received ; in original form}

\pagerange{\pageref{firstpage}--\pageref{lastpage}} \pubyear{}

\maketitle

\label{firstpage}

\begin{abstract}

Fundamental parameters and the carbon, nitrogen and oxygen abundances are determined for 22 B-type stars with distances $d \le 600$~pc and slow rotation ($v \sin i \le 66$~km~s$^{-1}$). The stars are selected according to their effective temperatures $T_{\rm eff}$ and surface gravities $\log g$, namely: $T_{\rm eff}$ is between 15300 and 24100~K and $\log g$ is mostly greater than 3.75; therefore, stars with medium masses of 5-11 M$_\odot$ are selected. Theory predicts for the stars with such parameters that the C, N and O abundances in their atmospheres should correspond to their initial values. Non-LTE analysis of C~II, N~II and O~II lines is implemented. The following mean C, N and O abundances are obtained: $\log\epsilon$(C)~=~8.31$\pm$0.13, $\log\epsilon$(N)~=~7.80$\pm$0.12 and $\log\epsilon$(O)~=~8.73$\pm$0.13. These values are in very good agreement with recent data on the C, N and O abundances for nearby B stars from other authors;
it is important that different techniques are applied by us and other authors. When excluding for the stars HR~1810 and HR~2938, which can be mixed, we obtain the following mean abundances for the remaining 20 stars: $\log\epsilon$(C)~=~8.33$\pm$0.11, $\log\epsilon$(N)~=~7.78$\pm$0.09 and $\log\epsilon$(O)~=~8.72$\pm$0.12; these values are in excellent agreement with a present-day Cosmic Abundance Standard (CAS) of Nieva \& Przybilla.

      The derived mean N and O abundances in unevolved B stars are very close to the solar photospheric abundances, as well as to the protosolar ones. However, the mean C abundance is somewhat lower than the solar one; this small but stable carbon deficiency is confirmed by other authors. One may suggest two possibilities to explain the observed C deficiency. First, current non-LTE computations of C~II lines are still partially inadequate. In this case the C deficiency is invalid, so one may conclude that the Sun and the local unevolved B stars have the same metallicity. This would mean that during the Sun's life (i.e., for the past $4.5\cdot10^9$ yr) the metallicity of the solar neighbourhood has not markedly changed; so, an intensive enrichment of the solar neighbourhood by metals occurred before the Sun's birth. Second, the C deficiency in the local B stars is valid; it is supposed that the Sun can migrate during its life from inner parts of the Galactic disk where it has born, so its observed chemical composition can differ from the composition of young stars in its present neighbourhood.

\end{abstract}

\begin{keywords}
stars: abundances - stars: evolution - stars: early-type
\end{keywords}

\section{Introduction}

Some recent works show that the metallicity of young stars in the solar neighborhood is very close to the Sun's metallicity. In particular, it was shown by Lyubimkov et al. (2010) that the mean iron abundance in the atmosphere of 48 A-, F- and G-type supergiants with distances $d < 700$~pc is $\log\epsilon$(Fe)~=~7.48$\pm$0.09, a value that coincides  with the solar abundance $\log\epsilon_\odot$(Fe)~=~7.50$\pm$0.04 (Asplund et al. 2009). Earlier it was found for local early B-type stars, the progenitors of AFG supergiants, that their mean magnesium abundance is $\log\epsilon$(Mg)~=~7.59$\pm$0.15 (Lyubimkov et al. 2005), that is  equal to the solar abundance $\log\epsilon_\odot$(Mg)~=~7.60$\pm$0.04 (Asplund et al. 2009). One may mention also the Fe abundance obtained by Luck, Kovtyukh \& Andrievsky (2006) for Cepheids in the solar vicinity, as well as the Fe and Mg abundances found by Fuhrmann (2004) for nearby F, G and K dwarfs and subgiants of the thin Galactic disk. Moreover, it was found for B-type stars in the Orion OB1 association that their S and O abundances are very close to the solar ones; see Daflon et al. (2009) and Sim\'{o}n-D\'{i}az (2010), respectively. Nieva \& Sim\'{o}n-D\'{i}az (2011) showed that the B stars in the Orion OB1 association show good agreement with the Sun in their C, N, O. Mg, Si, and Fe abundances. 

Just three years ago, the problem of metallicity of young stars seemed far from simple. 
It was shown by several authors that the metal abundances in nearby B-type stars were somewhat 
less than in the Sun (see, e.g., Morel's 2009 review). In particular, the derived C, N and O 
abundances of B-type stars tended to be less than in the Sun.   Two possible explanations were
considered, namely: (i) there is a real slight deficiency of C, N and O in the stellar atmospheres, 
e.g.,  connected with stellar evolution or (ii) the C, N and O abundances were systematically 
underestimated in some studies. 
Then in 2003-2008, studies were reported where 
the N and O abundances in B stars were close to the revised N and O solar abundances based on a 
three-dimensional model of the Sun's atmosphere; the C abundance continued to show a small 
deficiency (see, e.g., Herrero 2003; Herrero \& Lennon 2004 and Przybilla 2008).  
In order to investigate the possible metallicity difference between the Sun and young stars,
 we undertook an accurate analysis of the 
C, N and O abundances for a rather large sample of nearby early and medium B-type stars 
in the main sequence (MS) evolutionary phase.
(Note that up-to-date results for C, N and O appeared during  our analyses will 
be discussed in Section 7).

The C, N and O abundances in early and medium B-type stars are determined 
here from C~II, N~II and O~II spectral lines. These lines are rather sensitive 
to two fundamental parameters of the stars, namely their effective temperatures $T_{\rm eff}$ and 
surface gravities $\log g$. Recently, we showed (Lyubimkov et al. 2009, 2010) that a significant 
improvement in the accuracy of the $\log g$ values can be obtained through application of van 
Leeuwen's (2007) new reduction of the Hipparcos parallaxes. 
Using these new parallaxes in an analysis of A, F and G supergiants, we obtained $\log g$ with 
very high accuracy for stars with distances $d < 700$~pc. Now, we apply this method in the 
present study of B-type MS stars. 

      As far as the effective temperature $T_{\rm eff}$ is concerned, there is some difference in various $T_{\rm eff}$ scales for B stars that is especially significant for the hottest B stars with $T_{\rm eff} \ge  25000$~K. Therefore, we consider relatively cool B stars with $T_{\rm eff} \le 24000$~K, for which the $T_{\rm eff}$ estimates are more reliable. Moreover, we consider mostly those B stars with $\log g \ge 3.75$, i.e., stars rather far from the termination of the MS phase, in order to exclude possible evolutionary changes in the C, N and O abundances. In other words, we try to determine the $initial$ abundances of these elements in such young stars. 

      Carbon, nitrogen and oxygen participate in the CNO-cycle, which is the main source of energy of the B-type MS stars. The C and N abundances change significantly in stellar interiors during the MS phase (the O abundance shows smaller alterations). It has been predicted that the C, N and O abundances can change markedly in surface layers too, if  rotationally-induced mixing in the MS phase occurs (Heger \& Langer, 2000; Frischknecht et al. 2010). The changes in the surface C, N and O abundances by the MS phase termination depend strongly on the mass $M$ and the initial rotational velocity $v_0$ of a star: the greater $M$ and $v_0$ the stronger the predicted final alterations. Since our goal is to obtain the initial abundances, we consider the B-type stars with medium masses (5-11 M$_\odot$) that are not close to the MS phase termination. Their observed rotational velocities are rather small, too ($v \sin i \le 66$~km~s$^{-1}$) which implies that the majority of the selected
stars are slow rotators but an admixture of rapid rotators observed at high angles of
inclination may be present.

\section{SELECTION OF STARS}

We have studied previously a large sample of early and medium B-type stars in the MS phase; results have been published in four papers (Lyubimkov et al. 2000, 2002, 2004 and 2005; hereinafter Papers I, II, III and IV, respectively). In particular, in Paper III we determined from He~I lines the helium abundance $He/H$ and the projected rotational velocity $v \sin i$ for 102 B stars. The $v \sin i$ values were found to range from 5 to 280 km~s$^{-1}$; such a large variation allowed us to analyze a relation between the helium abundances and rotational velocities. The used He~I lines are strong in spectra of early and medium B stars, so their equivalent widths can be measured without problems even for the stars with rather high $v \sin i$. On the contrary, the C II, N II and O II lines used in the C, N and O abundance analysis are significantly weaker, so measurements of their equivalent widths for $v \sin i > 100$~km~s$^{-1}$ are difficult, especially for stars with effective temperatures $T_{\rm eff} < 17000$~K. As a result, we found that only 54 stars with $v \sin i < 100$ km~s$^{-1}$ from our original list are suitable for the C, N and O abundance analysis. 

      There are other limitations as well. One concerns the effective temperatures $T_{\rm eff}$ of B stars. It is known that the $T_{\rm eff}$ scales of various authors for B stars are different, and the difference increases with $T_{\rm eff}$, so the differences are especially large for the  hottest B stars with $T_{\rm eff} \ge 25000$~K (see, e.g., a comparison of some $T_{\rm eff}$ scales in Daflon, Cunha \& Butler 2004). We determined in Paper II the $T_{\rm eff}$ and $\log g$ values for more than 100 B stars and confirmed that uncertainties in the $T_{\rm eff}$ are especially great for $T_{\rm eff} \ge 25000$~K. Therefore, we selected from our original list only stars with $T_{\rm eff} \le 24000$~K.  The next limitation regards another fundamental parameter, the surface gravity $\log g$. In order to exclude as far as possible evolutionary effects on the observed C, N and O abundances, we consider stars which are not close to the MS phase termination. Therefore, the selected stars have mostly $\log g > 3.75$. 

      One more limitation regards the stellar parallaxes $\pi$, which are used by us as the  principal indicator of the surface gravity $\log g$. Using van Leeuwen's (2007) catalog, we selected stars with $\pi \ge 1.67$~mas, i.e., with distances $d \le 600$~pc (mas~=~milliarcsecond). Selection of stars with smaller $\pi$ values would reduce markedly the accuracy of the derived surface gravities $\log g$. 

\begin{table*}
\centering
\caption{Basic parameters of programme B stars}
\begin{tabular}{cccccccc}
\hline
HR  &   HD &   $\pi$, & $d$, & $v \sin i$, & $T_{\rm eff}$ & $\log g$ &  $M$/M$_\odot$ \\
    &      &     mas  & pc   & km~s$^{-1}$ & K&  cgs&  \\
\hline
38   &  829  & 1.81$\pm$0.42 & 552 $\pm$98 & 11.5$\pm$1.5 & 19220$\pm$510 & 3.83$\pm$0.15 & 7.6 \\
1617 & 32249 & 4.42$\pm$0.23 & 226 $\pm$11 & 45.0$\pm$1.0 & 18570$\pm$20  & 3.87$\pm$0.04 & 7.1 \\
1640 & 32612 & 1.76$\pm$0.41 & 568$\pm$102 & 53.5$\pm$1.5 & 19750$\pm$230 & 3.81$\pm$0.15 & 8.0 \\
1781 & 35299 & 3.72$\pm$0.32 & 269 $\pm$21 & 5.0 $\pm$2.5 & 23700$\pm$410 & 4.27$\pm$0.07 & 8.9 \\
1810 & 35708 & 5.20$\pm$0.21 & 192 $\pm$ 7 & 24.0$\pm$2.0 & 21040$\pm$520 & 4.11$\pm$0.07 & 7.7 \\
1820 & 35912 & 2.53$\pm$0.55 & 395 $\pm$67 & 13.5$\pm$2.0 & 19370$\pm$380 & 4.05$\pm$0.19 & 7.0 \\
1923 & 37356 & 2.03$\pm$0.47 & 493 $\pm$88 & 17.5$\pm$3.0 & 21400$\pm$630 & 3.78$\pm$0.16 & 9.5 \\
1933 & 37481 & 2.44$\pm$0.39 & 410 $\pm$55 & 66.0$\pm$2.5 & 24090$\pm$850 & 4.14$\pm$0.13 & 9.7 \\
2058 & 39777 & 2.80$\pm$0.53 & 357 $\pm$55 & 21.5$\pm$2.0 & 21720$\pm$370 & 4.24$\pm$0.16 & 7.7 \\
2205 & 42690 & 2.73$\pm$0.28 & 366 $\pm$34 & 9.8 $\pm$3.5 & 19390$\pm$310 & 3.74$\pm$0.04 & 8.1 \\
2344 & 45546 & 3.00$\pm$0.24 & 333 $\pm$25 & 61.0$\pm$2.5 & 19420$\pm$570 & 3.76$\pm$0.07 & 8.0 \\
2756 & 56342 & 5.19$\pm$0.22 & 193 $\pm$ 8 & 26.0$\pm$1.5 & 16590$\pm$390 & 4.06$\pm$0.04 & 5.4 \\
2824 & 58325 & 1.67$\pm$0.36 & 599$\pm$101 & 12.5$\pm$2.5 & 20210$\pm$450 & 3.86$\pm$0.16 & 8.4 \\
2928 & 61068 & 1.93$\pm$0.28 & 518 $\pm$64 & 26.0$\pm$1.5 & 23650$\pm$850 & 3.83$\pm$0.11 & 11.2\\
3023 & 63271 & 1.99$\pm$0.33 & 503 $\pm$69 & 42.0$\pm$1.5 & 22400$\pm$200 & 3.86$\pm$0.12 & 9.7 \\
7426 & 184171& 3.80$\pm$0.16 & 263 $\pm$11 & 29.5$\pm$2.0 & 16540$\pm$850 & 3.60$\pm$0.07 & 6.8 \\
7862 & 196035& 3.56$\pm$0.46 & 281 $\pm$32 & 34.0$\pm$1.5 & 17750$\pm$550 & 4.19$\pm$0.09 & 5.7 \\
7996 & 198820& 1.87$\pm$0.33 & 535 $\pm$78 & 35.0$\pm$1.5 & 15980$\pm$400 & 3.60$\pm$0.12 & 6.5 \\
8385 & 209008& 3.07$\pm$0.42 & 326 $\pm$38 & 19.0$\pm$3.0 & 15360$\pm$240 & 3.76$\pm$0.10 & 5.6 \\
8549 & 212883& 2.17$\pm$0.63 & 461 $\pm$95 & 8.0 $\pm$3.0 & 20400$\pm$680 & 3.97$\pm$0.21 & 7.8 \\
8768 & 217811& 3.07$\pm$0.58 & 326 $\pm$50 & 8.0 $\pm$4.0 & 18090$\pm$730 & 3.94$\pm$0.16 & 6.5 \\
9005 & 223128& 2.19$\pm$0.27 & 457 $\pm$49 & 15.5$\pm$2.5 & 22340$\pm$870 & 3.79$\pm$0.11 & 10.5\\
\hline
\end{tabular}
\end{table*}

      The final list of B stars selected is presented in Table~1; it contains 22 objects in all. We present there for each star its HR and HD numbers, parallax $\pi$ with its error from van Leeuwen's (2007) catalog and corresponding distance $d~=~1/\pi$, where $\pi$ is in arcseconds. The presented errors in distances $d$ are evaluated with the formula given by van Leeuwen's (2007, p.86). The projected rotational velocities $v \sin i$ with their uncertainties from Paper III are provided, too; note that the $v \sin i$ values have been found in Paper III from profiles of six He~I lines. Other parameters in Table~1, namely the effective temperature $T_{\rm eff}$, surface gravity $\log g$ and mass $M$, will be determined below. 

      The masses $M$ of the selected stars are between 5 and 11 M$_\odot$, and their projected rotational velocities are $v \sin i \le 66$~km~s$^{-1}$. Therefore, we may hope that possible evolutionary changes to the surface N and O abundances due to the rotationally-induced mixing in the MS phase are minimized. However, we need to be confident that the derived N and O abundances of these stars are really close to their $initial$ abundances; in other words, they have not been altered by mixing. In this connection, the stars with the lowest surface gravities $\log g$ are of special interest, because they are closer to the MS phase termination than other programme stars and, therefore, may show somewhat altered C, N and O abundances. There are seven stars in Table~1 with $\log g < 3.80$, specifically with $\log g$~=~3.60-3.79. Their masses are $M$~=~5.6-10.5 M$_\odot$; the projected rotational velocities are $v \sin i$~=~9-35 km~s$^{-1}$ for six stars and $v \sin i$~=~61 km~s$^{-1}$ for HR~2344. It follows from Frischknecht et al.'s (2010) computations that for the model with $M$~=~12 M$_\odot$ and initial rotational velocity $v_0$~=~62 km~s$^{-1}$ the surface value $\log\epsilon$(C) decreases only by 0.02 dex and $\log\epsilon$(N) increases only by 0.05 dex by the MS end. Therefore, for the lower masses and rotational velocities, like that for 7 stars in question, the expected changes in $\log\epsilon$(C) and $\log\epsilon$(N) are still smaller, whereas the changes in $\log\epsilon$(O) are imperceptible (according to the theory, oxygen shows smaller changes than carbon and nitrogen. So, the derived $\log\epsilon$(C), $\log\epsilon$(N) and $\log\epsilon$(O) values for all programme stars are anticipated to be very close to the initial abundances.

\section{OBSERVATIONS AND EQUIVALENT WIDTHS}

We used here the same high-resolution spectra as in Papers I-IV. The spectra were acquired in 1996-1998 at two observatories, namely the McDonald Observatory (McDO) of the University of Texas and the Crimean Astrophysical Observatory (CrAO). At the McDO the 2.7-m telescope and the Tull coud\'{e} echelle spectrometer was employed (Tull et al. 1995). The resolving power was $R = 60 000$ and the typical signal-to-noise ratio was between 100 and 300. At the CrAO we observed on the 2.6-m telescope with the coud\'{e} spectrograph. In this case we had $R = 30 000$ and a signal-to-noise ratio between 50 and 200.  A more detailed description of the observations and reductions of spectra was presented in Paper I. It should be noted that spectral intervals of about 70~\AA\ centered on He~I lines were recorded at the CrAO. Since these intervals contain few C~II, N~II and O~II lines, we used mostly the McDO echelle spectra for selection of the lines and measurements of their equivalent widths. 

      When measuring the equivalent widths $W$, we found that rarely could the C~II, N~II and O~II lines be accurately fitted by Gaussian profiles. Therefore, in most cases the $W$ values were measured by direct integration. We checked for possible blending of the lines using the database VALD (Kupka et al. 1999, Heiter et al. 2008) and predicted $W$ values for models based on parameters $T_{\rm eff}$ and $\log g$ from Paper II. Only those lines were selected where the estimated blending was markedly less than 10 per cent. It should be noted that the number of measured lines is different for different stars; this number depends on the effective temperature $T_{\rm eff}$ and the observed rotational velocity $v \sin i$. The lines in question are especially weak for stars with $T_{\rm eff} < 17000$~K, so in this case the $W$ values are measurable only for stars with low $v \sin i$. For some stars it was possible to obtain $W$ only for a few lines including the relatively strong lines C~II 4267~\AA\ and N~II 3995~\AA. We found that use of these lines in the C and N abundance determination is questionable, so they were excluded from the analysis (see below).

\begin{figure}
\begin{center}
\includegraphics[width=84mm]{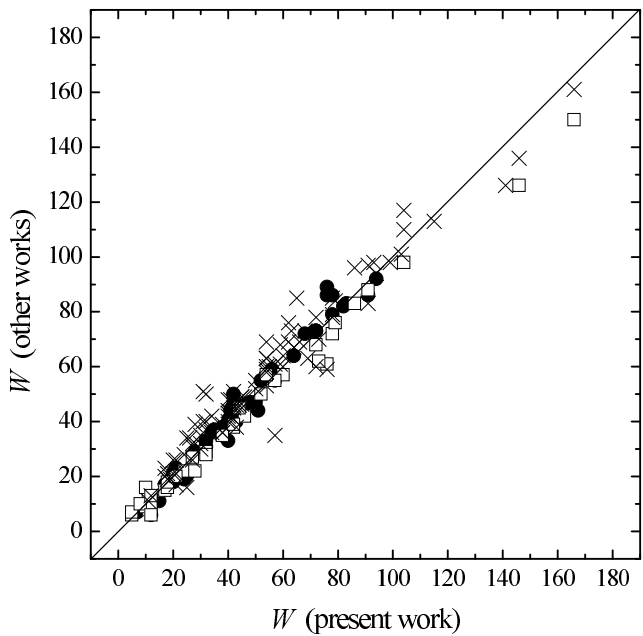}
\end{center}
\caption[]{Comparison of our equivalent widths (in m\AA) of N~II and O~II lines with data from other works, namely: Gies \& Lambert (1992) - squares, Cunha \& Lambert (1992, 1994) - filled circles, and Kilian \& Nissen (1989) - crosses. Straight line corresponds to the equal W values.}
\end{figure}

      It is interesting to compare the measured equivalent widths $W$ with other measurements. In Fig.~1 we compare our $W$ values with the data from three sources, namely Gies \& Lambert (1992), Cunha \& Lambert (1992, 1994) and Kilian \& Nissen (1989), where there are 3, 4 and 4 stars in common, respectively. One may see that there is good agreement for lines with $W < 120$~m\AA, and only the few lines with $W > 120$~m\AA\ show a small systematic discrepancy. So, one may conclude that there is, in general, good agreement   between our and previous $W$ measurements.

\section{NON-LTE COMPUTATIONS}

\subsection{Necessity for a non-LTE approach}

Departures from local thermodynamic equilibrium (LTE) have an influence on computations of 
C~II, N~II and O~II lines for early B stars. Using both LTE and non-LTE approach, Gies \& Lambert (1992) found that differences in the N and O abundances between these two cases are rather small on average, but for some stars they exceed 0.2 dex. An appreciable difference due to the non-LTE effects can appear for the derived microturbulent parameter $V_t$. According to Korotin et al. (1999b), non-LTE corrections to the oxygen abundance in early B stars can attain 0.2-0.3 dex. 

\begin{figure}
\begin{center}
\includegraphics[width=84mm]{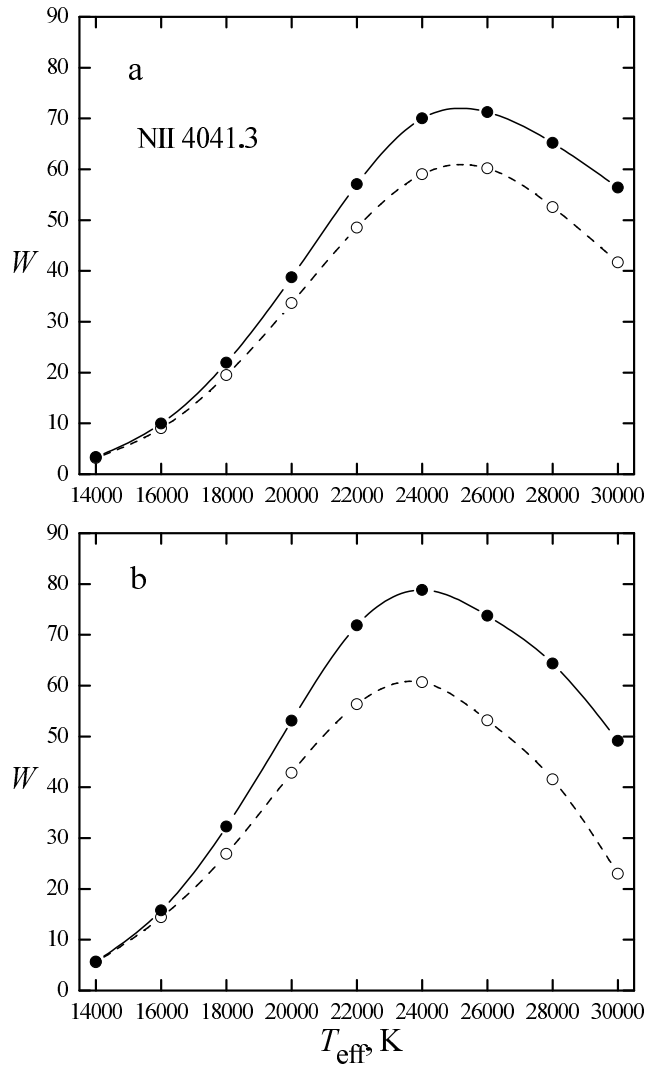}
\end{center}
\caption[]{Comparison of computed equivalent widths of the N~II 4041.3 \AA\ line for non-LTE (solid curve) and LTE (dashed curve). Computations are implemented for (a) $\log g = 4.0$ and (b) $\log g = 3.5$.}
\end{figure}

\begin{figure}
\begin{center}
\includegraphics[width=84mm]{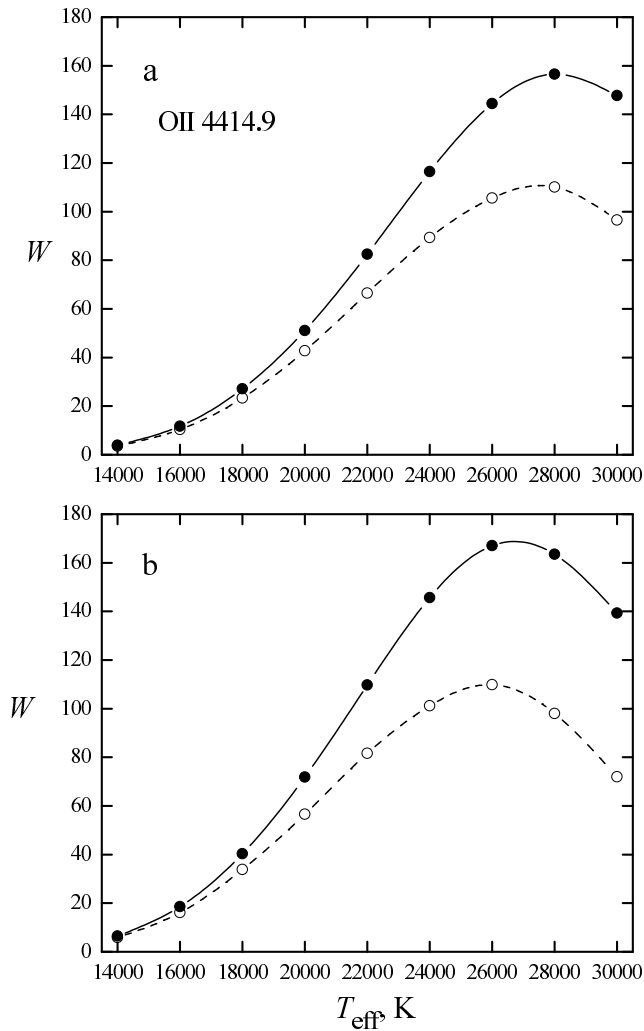}
\end{center}
\caption[]{Comparison of computed equivalent widths of the O~II 4414.9 \AA\ line for non-LTE (solid curve) and LTE (dashed curve). Computations are implemented for (a) $\log g = 4.0$ and (b) $\log g = 3.5$.}
\end{figure}

      In order to study the non-LTE effect on C~II, N~II and O~II lines in more detail, we performed both LTE and non-LTE calculations for all the lines used in our analysis - see the next section for details. We show as an example in Fig.~2 and Fig.~3 results of the calculations for two lines, namely N~II 4041.3~\AA\ and O~II 4414.9~\AA\ (the solar N and O abundances are adopted). Equivalent widths $W$ of the lines are displayed as a function of the effective temperature $T_{\rm eff}$ for (a) $\log g$~=~4.0 and (b) $\log g$~=~3.5. For these lines, as well as for other used lines, the $W$ values in the non-LTE case are greater than in the LTE one. The difference is small for stars with $T_{\rm eff} < 18000$~K but increases for hotter stars. Fig.~2 shows that in the $W$ maximum region of the line N~II 4041.3~\AA, i.e. at $T_{\rm eff}$~=~25000~K (a) or 24000~K (b), the difference is 18 per cent for $\log g$~=~4.0 and 30 per cent for $\log g$~=~3.5. According to Fig.~3,  the discrepancy between non-LTE and LTE for the O~II 4414.9~\AA\ line is more significant. In particular, in the $W$ maximum region at $T_{\rm eff}$~=~28000~K (a) or 26000~K (b) the difference is about 40 per cent for $\log g$~=~4.0 and 55 per cent for $\log g$~=~3.5. So, it follows from our computations that departures from LTE for B stars with $T_{\rm eff} > 20000$~K are substantial. Therefore, we use a non-LTE approach for all lines and all stars in the present work. 

      Our analysis showed that the non-LTE influence on derived abundances leads to a significant decrease of the microturbulent parameter $V_t$, especially for relatively hot B stars with $T_{\rm eff} > 20000$~K. The non-LTE $V_t$ values for such stars can decrease by a factor of about two as compared with the LTE ones. Therefore, we confirmed Gies \& Lambert's (1992) conclusion about the strong dependence of the $V_t$ derivation on the LTE or non-LTE approach.

\subsection{Non-LTE computations and selection of lines}


\tabcolsep=0.35cm
\begin{table*}
\caption{List of the used C~II, N~II and O~II lines}
\centering
\begin{tabular}{llrllrllr}

\hline
CII    & $E_l$, & $\log gf$ & NII    & $E_l$, & $\log gf$ & OII    & $E_l$, & $\log gf$ \\
$\lambda$,~\AA & eV   &           & $\lambda$,~\AA & eV   &           & $\lambda$,~\AA & eV   & \\
\hline
3918.969 & 16.332 & -0.533  &  4035.080 & 23.125 &  0.623  &  4072.150 & 25.650 &  0.552 \\
3920.681 & 16.333 & -0.232  &  4041.310 & 23.142 &  0.853  &  4075.859 & 25.665 &  0.692 \\
5137.253 & 20.702 & -0.940  &  4043.533 & 23.132 &  0.743  &  4085.114 & 25.650 & -0.188 \\
5139.175 & 20.704 & -0.740  &  4199.977 & 23.246 &  0.030  &  4132.800 & 25.832 & -0.066 \\
5143.495 & 20.704 & -0.212  &  4227.738 & 21.600 & -0.068  &  4156.533 & 25.849 & -0.696 \\
5145.165 & 20.710 &  0.189  &  4447.028 & 20.409 &  0.285  &  4185.451 & 28.358 &  0.604 \\
5151.083 & 20.710 & -0.179  &  4601.478 & 18.466 & -0.428  &  4303.821 & 28.822 &  0.640 \\
5648.070 & 20.704 & -0.424  &  4607.154 & 18.462 & -0.507  &  4317.134 & 22.966 & -0.385 \\
5662.456 & 20.710 & -0.249  &  4613.870 & 18.466 & -0.665  &  4395.939 & 26.249 & -0.169 \\
6578.048 & 14.449 & -0.026  &  4621.394 & 18.466 & -0.514  &  4414.889 & 23.442 &  0.171 \\
6582.873 & 14.449 & -0.328  &  4630.535 & 18.483 &  0.094  &  4452.342 & 23.442 & -0.789 \\
6779.940 & 20.704 &  0.024  &  4643.083 & 18.483 & -0.359  &  4590.978 & 25.661 &  0.350 \\
6783.904 & 20.710 &  0.304  &  4678.136 & 23.572 &  0.434  &  4610.197 & 29.063 & -0.170 \\
         &        &         &  4987.377 & 20.940 & -0.555  &  4638.854 & 22.966 & -0.332 \\
         &        &         &  5002.700 & 18.462 & -1.021  &  4641.813 & 22.980 &  0.054 \\
         &        &         &  5005.147 & 20.666 &  0.592  &  4649.139 & 22.999 &  0.308 \\
         &        &         &  5007.330 & 20.940 &  0.171  &  4650.839 & 22.966 & -0.361 \\
         &        &         &  5010.619 & 18.466 & -0.606  &  4661.632 & 22.980 & -0.277 \\
         &        &         &  5025.654 & 20.666 & -0.546  &  4673.735 & 22.980 & -1.088 \\
         &        &         &  5045.092 & 18.483 & -0.407  &  4676.237 & 22.999 & -0.395 \\
         &        &         &  5666.628 & 18.466 & -0.045  &  4701.180 & 28.830 &  0.088 \\
         &        &         &  5679.554 & 18.483 &  0.250  &  4703.163 & 28.513 &  0.262 \\
         &        &         &  5686.213 & 18.466 & -0.549  &  4890.865 & 26.305 & -0.436 \\
         &        &         &  6482.043 & 18.497 & -0.162  &  4906.841 & 26.305 & -0.160 \\
         &        &         &           &        &         &  4941.064 & 26.554 & -0.054 \\
         &        &         &           &        &         &  4943.000 & 26.561 &  0.239 \\
\hline
\end{tabular}
\end{table*}

For the non-LTE computations of C~II, N~II and O~II lines, we applied the code MULTI created by Carlsson (1986) and updated by S.A.Korotin (see Korotin et al. 1998, 1999 a,b). This code, as well as all input data including the N and O model atoms was kindly provided to us by S.A. Korotin (the carbon model atom is taken from Sigut, 1996). Note that we used this same code in a non-LTE analysis of the nitrogen abundance from N~I lines for A- and F-type supergiants (Lyubimkov et al. 2011). Atomic data for C~II, N~II and O~II lines are taken from the database VALD (Kupka et al. 1999; Heiter et al. 2008). The list of the lines used is presented in Table 2 including their lower excitation potentials $E_l$ and the oscillator strengths $\log gf$. In the computations we used, as in Paper III and IV, the plane-parallel blanketed LTE model atmospheres, which were computed by us with the code ATLAS9 (Kurucz 1993) for the corresponding parameters $T_{\rm eff}$ and $\log g$ of programme stars.  Our approach is what has been termed a hybrid method, i.e., LTE model atmospheres are combined with non-LTE line formation.

      As mentioned above, 22 B stars with effective temperatures $T_{\rm eff} \le 24000$~K were selected for the present work. However, we plan to analyze later the C, N and O abundances for hotter B stars from our original list with $T_{\rm eff}$ up to 30000~K. Therefore, from the outset, we performed non-LTE computations of C~II, N~II and O~II lines for  B stars with effective temperatures $T_{\rm eff}$ up to 30000~K and rotational velocities $v \sin i < 100$~km~s$^{-1}$. As a result, we considered 54 objects in all instead of 22. Their parameters $T_{\rm eff}$ and $\log g$ were taken from Paper II as a first approximation. 

      We compared for each of the 54 stars the non-LTE C, N and O abundances derived from individual C~II, N~II and O~II lines with the mean C, N and O abundances obtained from other lines. 
We found that some lines display a systematic discrepancy. In particular, the strongest C~II line in the visible spectral region, namely the line C~II 4267~\AA, showed a systematic difference in $\log\epsilon$(C) for hotter stars of our sample: for stars with $T_{\rm eff}$ between 20000 and 24000~K the $\log\epsilon$(C) underestimation from 4267~\AA\ is about 0.2-0.7 dex. A tendency of 4267~\AA\ to lower the $\log\epsilon$(C) values for early B stars has been noted earlier. Meantime, a few attempts have been made to update the non-LTE calculations of this line by using a more detailed model atom. In particular, Sigut's (1996) computations improved markedly the agreement with observed equivalent widths of the line. We applied Sigut's model atom in our non-LTE calculations of C~II lines. Nevertheless, it was impossible to eliminate the significant discrepancy between 4267~\AA\ and other C~II lines.

\begin{figure}
\begin{center}
\includegraphics[width=84mm, angle=-90, scale=0.83]{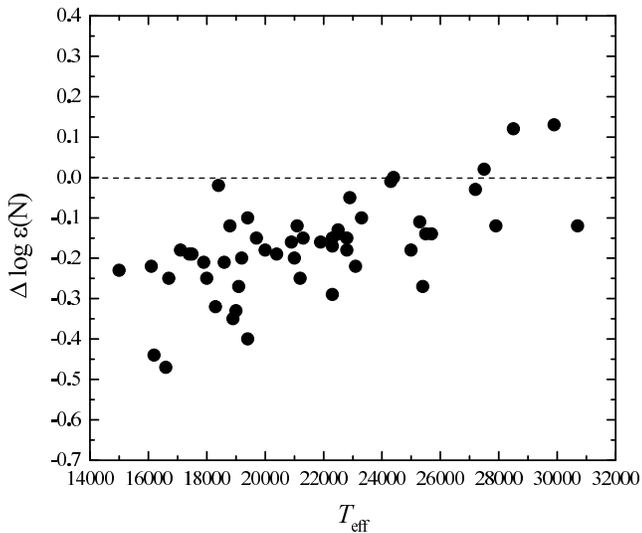}
\end{center}
\caption[]{Difference between the N abundance derived from the N II 3995 \AA\ line and the mean N abundance obtained from all other N II lines.}
\end{figure}

      Furthermore, a systematic difference was found  for the strongest N~II line in the visible spectral region: this line at N~II 3995~\AA\ is frequently used in analyses of B and O stars. In Fig.~4 the difference between the $\log\epsilon$(N) values determined from the N~II 3995~\AA\ line and the mean N abundances (without 3995~\AA) is shown as a function of $T_{\rm eff}$. 
The underestimate is obviously correlated with $T_{\rm eff}$ with the largest underestimation (up to 0.2-0.4 dex) occurring for the coolest B stars of the sample. If included in the collection of N~II lines, this line has a significant influence on the determination of the microturbulent parameter $V_t$. It is clear that the 3995~\AA\ line should be excluded from the analysis of our programme stars with $T_{\rm eff} \le 24000$~K. Nevertheless, as Fig.~4 shows, for the stars with $T_{\rm eff} > 26000$~K this line displays no systematic discrepancy; therefore, for such hot stars use of 3995~\AA\ in the N abundance analysis is likely to be permissible. 

      Thus, we excluded from further analyses the C II 4267~\AA\ and N~II 3995~\AA\ lines. Moreover, two O~II lines, namely 4093~\AA\ and 4705~\AA, were excluded because they show also a systematic trend with $T_{\rm eff}$.  It is interesting that in the recent work of Nieva \& Przybilla (2012), where updated model atoms for carbon and nitrogen were applied, the lines C~II 4267~\AA\ and N~II 3995~\AA\ did not show a marked difference from other lines. Therefore, the above-mentioned discrepancies for the lines C~II 4267~\AA\ and N~II 3995~\AA\ and two O~II lines are likely to be due to some incompleteness of the C, N and O model atoms used in our analysis (Korotin et al. 1998, 1999 a,b).

\section{EFFECTIVE TEMPERATURES $T\lowercase{_{\rm eff}}$ AND SURFACE GRAVITIES $\log\lowercase{g}$}

Effective temperatures $T_{\rm eff}$ and surface gravities $\log g$ of more than 100 B stars have been determined by us in Paper II. Now, we redetermine these parameters for the selected 22 stars using the new stellar parallaxes of van Leeuwen (2007) as principal indicators of $\log g$. Other indicators of $\log g$ used in Paper II, namely the $\beta$-index and equivalent widths of the Balmer lines H$_{\beta}$ and H$_{\gamma}$, are considered now only for comparison. As far as the effective temperature $T_{\rm eff}$ is concerned, we use the same two indicators as in Paper II, i.e., the colour indices $Q$ and $[c_1]$ in the photometric systems $UBV$ and $uvby$, respectively (as known, both indices are 
insensitive to interstellar extinction). 

      Observed $Q$ values were determined with photometric data from Mermilliod \& Mermilliod's (1994) and Mermilliod's (1994) catalogues. Observed $[c_1]$ and $\beta$ indices are found from Hauck \& Mermilliod's (1998) catalogue. Observed equivalent widths $W$(H$_{\beta}$) and $W$(H$_{\gamma}$) are provided in Paper I. These observed parameters are compared with the computed values. Computations of $W$(H$_{\beta}$) and $W$(H$_{\gamma}$) for a large number of model atmospheres have been presented by Kurucz (1993). Colour indices for calculations of $Q$ and $[c_1]$ are taken by us from Castelli \& Kurucz (2003). Computed $\beta$ values are published by Castelli \& Kurucz (2006). 

      As a typical example of the combined $T_{\rm eff}$ and $\log g$ determination, we show in Fig.~5 the diagram for the star HR~1810. This is the standard $T_{\rm eff}$ - $\log g$ diagram, where various curves are the loci from the theoretical models which predict the observed value for each of the considered parameters ($Q$, $[c_1]$, etc.). The locus that is set by parallax $\pi$ is virtually a straight line; it depends on $T_{\rm eff}$ very slightly. Note that application of $\pi$ was described in detail by us earlier (Lyubimkov et al. 2009, 2010). In the case of HR~1810 the parallax is rather large and known with a good accuracy, namely $\pi$~=~5.20$\pm$0.21 mas. When the effective temperature $T_{\rm eff}$ is fixed, the surface gravity of the star is derived reliably from $\pi$; in fact, the uncertainty in $\log g$ is only $\pm$0.03. 

      Two indicators of $T_{\rm eff}$, namely indices $Q$ and $[c_1]$, show in Fig.~5 somewhat different temperatures $T_{\rm eff}$; the difference is about 1000~K. We adopted the mean $T_{\rm eff}$ value, so the final parameters of the star HR~1810 are $T_{\rm eff}$~=~21040$\pm$520~K and $\log g$~=~4.11$\pm$0.07 (filled circle in Fig.~5). Fig.~5 shows that there is good agreement between the value $\log g$~=~4.11 derived from the parallax and the $\log g$ estimates from the Balmer lines and $\beta$-index. Using this technique, we obtained the parameters $T_{\rm eff}$ and $\log g$ for all programme stars; these values and their uncertainties are presented in Table~1. 

\begin{figure}
\begin{center}
\includegraphics[width=84mm]{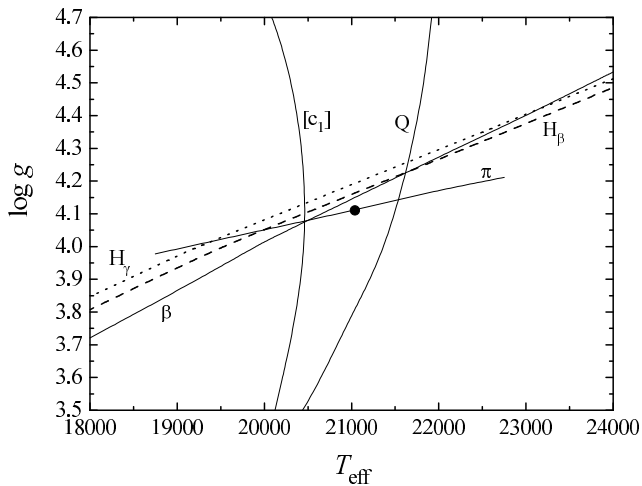}
\end{center}
\caption[]{The $T_{\rm eff}$ - $\log g$ diagram for the star HR~1810. The filled circle corresponds to the adopted parameters $T_{\rm eff} = 21040$~K and $\log g = 4.11$.}
\end{figure}

      Fig.~5 displays one feature in the $T_{\rm eff}$ determination for early B stars, which has been already noted by us in Paper~II, namely: there is a systematic discrepancy between indices $Q$ and $[c_1]$ as indicators of  the $T_{\rm eff}$.  An especially hotter $T_{\rm eff}$ from $Q$ as compared with $[c_1]$ is found for B stars with $T_{\rm eff} > 24000$~K, i.e., for spectral subtypes B0-B1.5. Our consideration of this discrepancy for B0-B1.5 stars in Paper~II led to the conclusion that the $T_{\rm eff}$($[c_1]$) values are more correct than $T_{\rm eff}$($Q$). Theoretical $Q$ values based on Kurucz's (1993) computed indices $(U-B)$ and (B-V) have been corrected in order to eliminate disagreement between $T_{\rm eff}$($Q$) and $T_{\rm eff}$($[c_1]$). Therefore, the index $[c_1]$ has been adopted as a principal indicator of $T_{\rm eff}$.  Now  that we use the computed  $(U-B)$ and $(B-V)$ indices from Castelli \& Kurucz (2003), we find that the discrepancy between $T_{\rm eff}$($Q$) and $T_{\rm eff}$($[c_1]$) decreases but does not disappear. 

      One may see from Table~1 that our programme B stars have effective temperatures from 15360 to 24090~K. For these relatively cool B stars the difference between $T_{\rm eff}$($Q$) and $T_{\rm eff}$($[c_1]$) leads to an  uncertainty in $T_{\rm eff}$ of between $\pm$20 and $\pm$870~K with a mean value of $\pm$500~K. When comparing such accuracy with data on $T_{\rm eff}$ from other authors, one may conclude that the accuracy is plenty good enough. 

      It is interesting to compare the $T_{\rm eff}$ and $\log g$ values from Table~1 with the old $T_{\rm eff}$ and $\log g$ values for the same 22 stars from Paper~II. This comparison is shown in Fig.~6. One sees that new effective temperatures $T_{\rm eff}$ are slightly hotter than the old ones (upper panel). The mean difference is about 3 per cent, but for three hottest programme stars with $T_{\rm eff}$ close to 24000~K it reaches 4-5 per cent. This $T_{\rm eff}$ increment is mainly explained by the fact that now we added the second independent $T_{\rm eff}$ indicator, namely the index $Q$, which tends to increase $T_{\rm eff}$ in comparison with $[c_1]$. 

\begin{figure}
\begin{center}
\includegraphics[width=84mm]{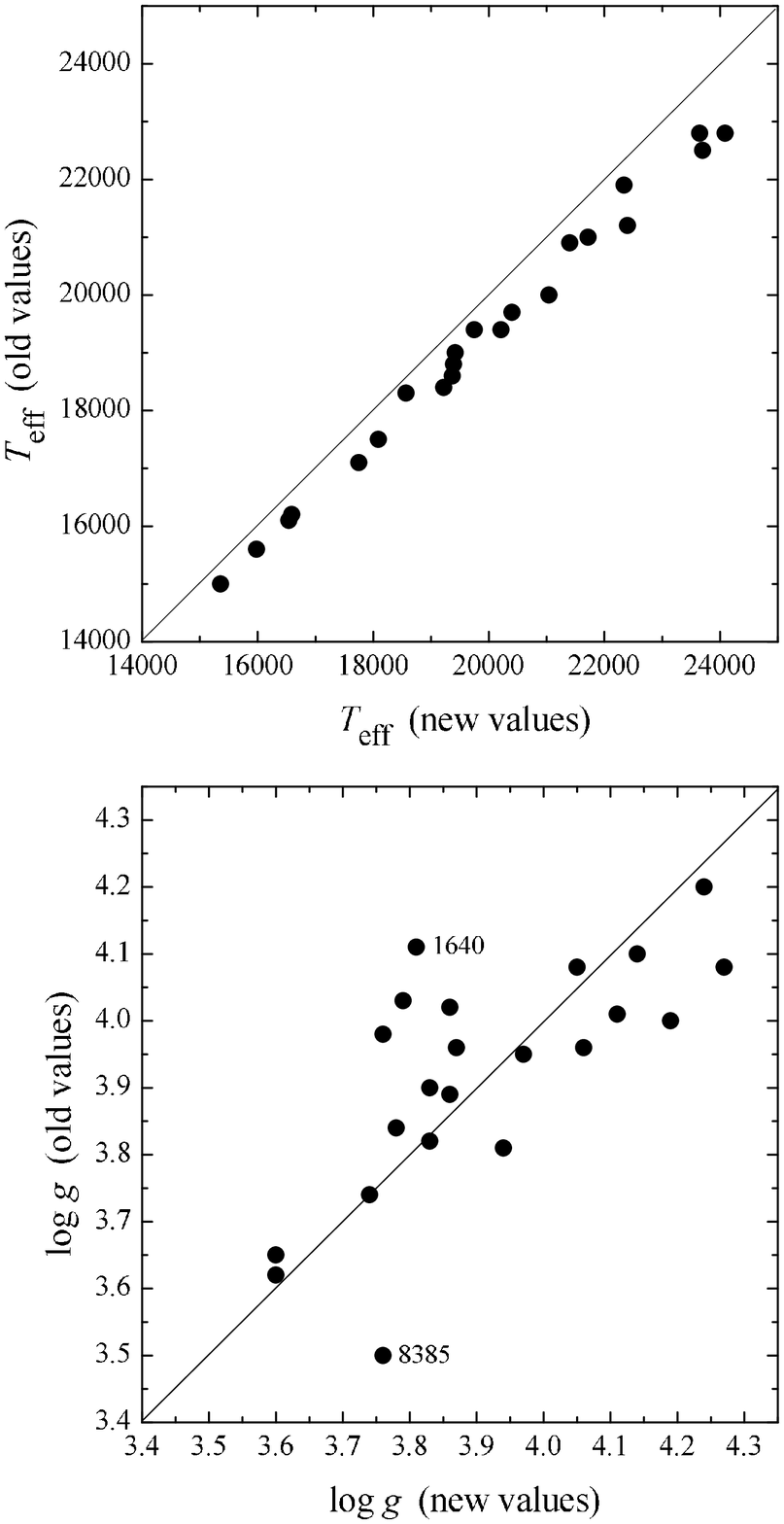}
\end{center}
\caption[]{Comparison of the new Teff and $\log g$ values with old parameters from Paper II.}
\end{figure}

      Lower panel of Fig.~6 presents the comparison of new and old $\log g$ values. It is important to emphasize that our new surface gravities are found from stellar parallaxes and, therefore, are independent of model atmospheres of the stars. On the contrary, when the old $\log g$ values in Paper~II were  determined, the $\beta$-index and equivalent widths of the Balmer lines H$_{\beta}$ and H$_{\gamma}$ were used; their observed values were compared with results of model computations. In other words, the new and old $\log g$ values are derived by different methods. One may see that there is a marked difference between the new and old $\log g$ values for some stars. For instance, the stars HR~1640 and HR~8385 (marked in Fig.~6) show a difference of +0.30 and -0.26 dex, respectively. Errors in their new $\log g$ values are $\pm$0.15 and $\pm$0.10 dex, respectively (Table~1), so they can explain the discrepancy only partially. We believe that the new surface gravities $\log g$ derived from parallaxes are more reliable. It should be noted that the mean error of the $\log g$ values in Table~1 is $\pm$0.11 dex, whereas an accuracy of $\pm$0.2 or $\pm$0.3 dex was considered typical for B stars a few years ago. Thus, our surface gravities $\log g$ derived from stellar parallaxes are rather accurate.

    What may we say about the  accuracy of the derived effective temperatures $T_{\rm eff}$? To answer this question, we compare our $T_{\rm eff}$ values with evaluations based on two other accurate $T_{\rm eff}$ scales. Napiwotzki et al. (1993) provided the $T_{\rm eff}$ calibration for $[u-b]$, $(b-y)_0$ and $(B-V)_0$ using a sample of stars with well known temperatures. Later Napiwotzki (2004) presented a new calibration, which is based on colors derived from Kurucz's (1993) model atmospheres. A scale provided by Fitzpatrick \& Massa (2005), which gives $T_{\rm eff}$ values in very good agreement with those derived with Napiwotzki's (2004) code,  is based on a new calibration of synthetic photometry in various photometric systems for 45 normal nearby B and early A stars on the basis of Kurucz's (1993) model atmospheres. Fitzpatrick \& Massa emphasize that their and Napiwotzki's calibrations 'have no input data in common'. In fact, these calibrations use different observables: Napiwotzki (2004) applies $uvby\beta$ photometry, whereas Fitzpatrick \& Massa (2005) employ IUE spectrophotometry, V-band fluxes and $Hipparcos$ parallaxes (old catalog, ESA 1997). 

\begin{figure}
\begin{center}
\includegraphics[width=84mm]{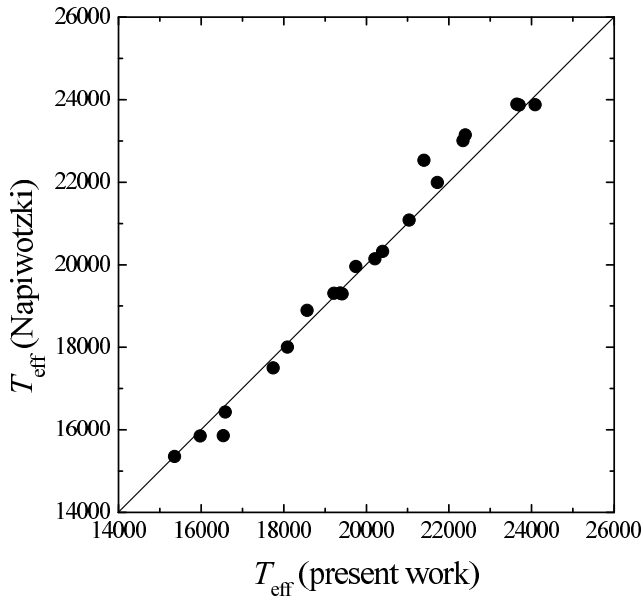}
\end{center}
\caption[]{Comparison of our $T_{\rm eff}$ values with ones derived with Napiwotzki's (2004) code. The straight line corresponds to equal effective temperatures.}
\end{figure}

      We show in Fig.~7 a comparison of our $T_{\rm eff}$ values from Table~1 with ones derived by us from $uvby\beta$-photometry (Hauck \& Mermilliod 1998) with Napiwotzki's (2004) code. One may see that there is very good agreement with Napiwotzki's $T_{\rm eff}$ scale and, therefore, with Fitzpatrick \& Massa's scale. Note especially that no systematic discrepancy is seen. This agreement with these two independent accurate $T_{\rm eff}$ scales confirms the   reliability of our new $T_{\rm eff}$ values. 

\tabcolsep=0.17cm
\begin{table}
\caption{Comparison of our parameters with those of Nieva \& Sim\'{o}n-D\'{i}az (2011) = NS'11 for two common stars, namely HR~1781 and HR~1820 (the NS'11 $\log\epsilon$(O) values are taken from Sim\'{o}n-D\'{i}az, 2010) }
\centering
\begin{tabular}{l@{}c@{}cc@{}c}
\hline
Parameter & \multicolumn{2}{c}{HR 1781 = HD 35299} & \multicolumn{2}{c}{HR 1820 = HD 35912}\\
                   &  Present work  &      NS'11     &   Present work  &     NS'11      \\
\hline
$T_{\rm eff}$, K   & 23700$\pm$410  & 24000$\pm$200  &  19370$\pm$380  & 19000$\pm$300  \\
$\log g$           &  4.27$\pm$0.07 &  4.20$\pm$0.08 &   4.05$\pm$0.19 &  4.00$\pm$0.10 \\
$V_t$, km~s$^{-1}$ &   1.2$\pm$1.0  &     0$\pm$1    &    0.8$\pm$ 1.0 &     2$\pm$1    \\
$\log\epsilon$(C)  &  8.25$\pm$0.10 &  8.37$\pm$0.07 &   8.28$\pm$0.10 &  8.33$\pm$0.09 \\
$\log\epsilon$(N)  &  7.76$\pm$0.08 &  7.81$\pm$0.07 &   7.76$\pm$0.09 &  7.76$\pm$0.07 \\
$\log\epsilon$(O)  &  8.68$\pm$0.07 &  8.71$\pm$0.08 &   8.71$\pm$0.08 &  8.77$\pm$0.12 \\
\hline
\end{tabular}
\end{table}

Additional checks on the parameters are possible through comparisons for stars common
to recent studies. For example,
Nieva \& Sim\'{o}n-D\'{i}az (2011) studied 13 early B-type stars in the Ori OB1 association including derivations of the C,  N and O abundances. Their sample includes two stars from Table~1, namely HR~1781 and HR~1820. We compare in Table~3 the parameters $T_{\rm eff}$ and $\log g$ of these stars from their and our work. One may see that there is very good agreement with the $T_{\rm eff}$ and $\log g$ values; the differences are within errors of the $T_{\rm eff}$ and $\log g$ determinations. It is important to note that Nieva \& Sim\'{o}n-D\'{i}az used quite different methods to us for their $T_{\rm eff}$ and $\log g$ determinations, i.e., their methods are based on  the non-LTE analysis of lines of C, O, Si and Fe in two stages of ionization. So, the comparison with their data confirms the reliability of our new $T_{\rm eff}$ and $\log g$ values. Nieva \& Sim\'{o}n-D\'{i}az adopt as we do the Kurucz ATLAS model atmospheres.

\tabcolsep=0.07cm
\begin{table}
\caption{Comparison of our parameters and the C, N and O abundances with those of 
Nieva \& Przybilla (2012) = NP'12 for four common stars}
\centering
\begin{tabular}{lccccc}
\hline

Parameter           & \multicolumn{2}{c}{HR 1781 = HD 35299}&\multicolumn{2}{c}{HR 1810 = HD 35708}\\
                    &   Present work  &   NP'12           &    Present work &    NP'12          \\
\hline                                                                       
$T_{\rm eff}$, K    &  23700$\pm$410  &  23500$\pm$300    &  21040$\pm$520  &  20700$\pm$200    \\
$\log g$            &   4.27$\pm$0.07 &   4.20$\pm$0.05   &   4.11$\pm$0.07 &   4.15$\pm$0.07   \\
$V_t$, km~s$^{-1}$  &    1.2$\pm$1.0  &      0$\pm$1      &    0.0$\pm$1.0  &      2$\pm$1      \\
$\log\epsilon$(C)   &   8.25$\pm$0.10 &   8.35$\pm$0.09   &   8.25$\pm$0.12 &   8.30$\pm$0.09   \\
$\log\epsilon$(N)   &   7.76$\pm$0.08 &   7.82$\pm$0.08   &   8.16$\pm$0.07 &   8.22$\pm$0.07   \\
$\log\epsilon$(O)   &   8.68$\pm$0.07 &   8.84$\pm$0.09   &   8.65$\pm$0.06 &   8.82$\pm$0.11   \\
\hline
                    &\multicolumn{2}{c}{HR 2928 = HD 61068}&\multicolumn{2}{c}{HR 8385 = HD 209008}\\
                    &   Present work  &   NP'12           &    Present work &    NP'12          \\
\hline                                                                       
$T_{\rm eff}$, K    &  23650$\pm$850  &  26300$\pm$300    &  15360$\pm$240  &   15800$\pm$200   \\
$\log g$            &   3.83$\pm$0.11 &   4.15$\pm$0.05   &   3.76$\pm$0.10 &    3.75$\pm$0.05  \\
$V_t$, km~s$^{-1}$  &    3.7$\pm$1.0  &      3$\pm$1      &    1.2$\pm$1.0  &       4$\pm$1     \\
$\log\epsilon$(C)   &   8.01$\pm$0.10 &   8.27$\pm$0.07   &   8.57$\pm$0.19 &    8.33$\pm$0.09  \\
$\log\epsilon$(N)   &   7.91$\pm$0.10 &   8.00$\pm$0.12   &   7.97$\pm$0.12 &    7.80$\pm$0.11  \\
$\log\epsilon$(O)   &   8.69$\pm$0.11 &   8.76$\pm$0.09   &   8.92$\pm$0.12 &    8.80$\pm$0.11  \\
\hline
\end{tabular}
\end{table}

      Nieva \& Przybilla's (2012) study of 29 early B stars contains four objects in common with our list. (Note that this 2012  paper determines $T_{\rm eff}$ and $\log g$ by the methods very similar to those used by Nieva \& Sim\'{o}n-D\'{i}az, 2011). According to Table 4, there is  good agreement between the parameters $T_{\rm eff}$ and $\log g$ of these stars and our values, except for HR~2928; for this star Nieva \& Przybilla (2012) obtained a markedly hotter effective temperature $T_{\rm eff}$ and, as a result, the greater $\log g$ value. It should be noted that Napiwotzki's (2004) method gives for HR~2928 the parameters $T_{\rm eff}$~= 23980~K and $\log g$~= 3.83 that are very close to ours.

      When basic parameters $T_{\rm eff}$ and $\log g$ are known, one may determine the mass $M$ of a star from evolutionary tracks. The derived masses $M$ of the  programme stars (in the solar masses M$_\odot$) are presented in Table~1; they are found from Claret's (2004) evolutionary tracks. One sees that the masses range from 5.4 to 11.2 M$_\odot$. The typical error in the $M$ determination due to uncertainties in $T_{\rm eff}$ and $\log g$ is $\pm$8 per cent.

      A question may arise about the reliability of stellar masses $M_{ev}$ derived from evolutionary tracks. It is known that the most reliable masses $M_{orb}$ are obtained for components of binaries from an analysis of their orbital elements. Lyubimkov (1996) compared the $M_{ev}$ and $M_{orb}$ values for a wide mass range; one may see from this comparison that there is very good agreement between $M_{ev}$ and $M_{orb}$ for masses between 5 and 15 M$_\odot$ (see Fig.~9 there). So, we believe that our $M$ values in the 5-11 M$_\odot$ range inferred from Claret's (2004) evolutionary tracks are rather accurate.

\begin{figure}
\begin{center}
\includegraphics[width=84mm]{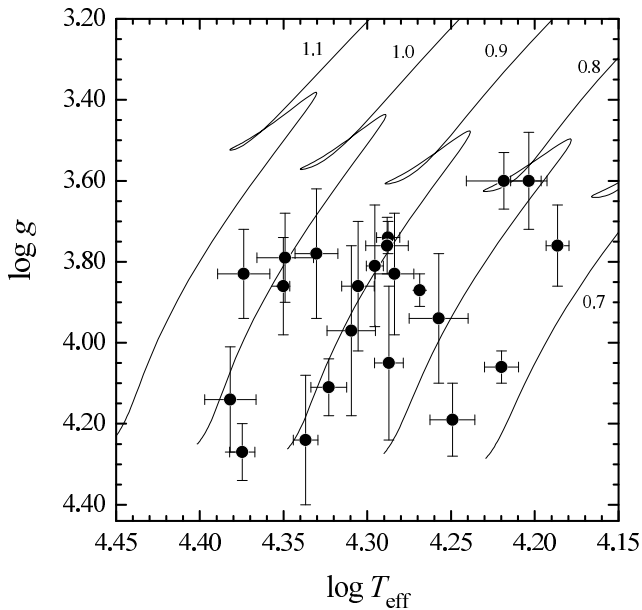}
\end{center}
\caption[]{Claret's (2004) evolutionary tracks and positions of programme stars on the $\log T_{\rm eff}$ -- $\log g$ plane. Error bars in $\log T_{\rm eff}$ and $\log g$ are shown for each star.}
\end{figure}

      In Fig.~8 we show Claret's (2004) tracks in the $T_{\rm eff}$--$\log g$ plane, as well as locations of all 22 stars with corresponding error bars for $T_{\rm eff}$ and $\log g$; note that the labels for the tracks are $\log M$/M$_\odot$. One  sees that all the stars are inside the MS band. Two stars, namely HR~7426 and HR~7996 with the lowest surface gravity $\log g$~= 3.60, are close to the MS phase termination. Nonetheless, as mentioned in Section 2, the observed C, N and O abundances for such stars (they have $M$ $\sim$ 6-7 M$_\odot$ and $v \sin i$ $\sim$ 30~km~s$^{-1}$) are expected to be close to the initial ones; rotationally-induced mixing is not predicted to affect surface abundances unless a slow rotator is a rapid rotator seen at a high angle of inclination.

\section{CARBON, NITROGEN AND OXYGEN ABUNDANCES}

\tabcolsep=0.10cm
\begin{table}
\caption{Carbon, nitrogen and oxygen abundances in atmospheres of 22 B-type MS stars. The number of lines used is shown in brackets.}
\begin{minipage}{84mm}
\centering
\begin{tabular}{lllll}
\hline
  HR & $V_t$ & $\log\epsilon$(C)  & $\log\epsilon$(N)   &  $\log\epsilon$(O) \\
     & km~s$^{-1}$ &              & \\
\hline
  38 &  2.7  & 8.29$\pm$0.09 (9)  &  7.70$\pm$0.06 (11) &  8.56$\pm$0.12 (8) \\
1617 &  0.5* & 8.16$\pm$0.06 (6)  &  7.63$\pm$0.11 (11) &  8.63$\pm$0.15 (11)\\
1640 &  0.0  & 8.36$\pm$0.13 (4)  &  7.78$\pm$0.13 (11) &  8.83$\pm$0.11 (9) \\
1781 &  1.2  & 8.25$\pm$0.10 (11) &  7.76$\pm$0.08 (23) &  8.68$\pm$0.07 (16)\\
1810 &  0.0  & 8.25$\pm$0.12 (9)  &  8.16$\pm$0.07 (13) &  8.65$\pm$0.06 (8) \\
1820 &  0.8  & 8.28$\pm$0.10 (6)  &  7.76$\pm$0.09 (19) &  8.71$\pm$0.08 (12)\\
1923 &  1.1  & 8.25$\pm$0.06 (9)  &  7.74$\pm$0.08 (20) &  8.62$\pm$0.06 (12)\\
1933 &  1.8  & 8.34$\pm$0.14 (4)  &  7.81$\pm$0.12 (12) &  8.58$\pm$0.10 (6) \\
2058 &  0.8  & 8.25$\pm$0.09 (9)  &  7.72$\pm$0.10 (22) &  8.68$\pm$0.09 (16)\\\
2205 &  1.9  & 8.33$\pm$0.09 (8)  &  7.76$\pm$0.07 (20) &  8.66$\pm$0.07 (17)\\
2344 &  0.0* & 8.21$\pm$0.02 (3)  &  7.88$\pm$0.20 (5)  &  8.73$\pm$0.08 (5) \\
2756 &  0.0* & 8.46$\pm$0.19 (5)  &  7.82$\pm$0.03 (4)  &  8.91$\pm$0.07 (4) \\
2824 &  1.0  & 8.23$\pm$0.10 (8)  &  7.66$\pm$0.09 (22) &  8.63$\pm$0.12 (20)\\
2928 &  3.7  & 8.01$\pm$0.10 (7)  &  7.91$\pm$0.10 (19) &  8.69$\pm$0.11 (16)\\
3023 &  1.2  & 8.30$\pm$0.12 (9)  &  7.79$\pm$0.13 (16) &  8.68$\pm$0.11 (12)\\
7426 &  0.0* & 8.30$\pm$0.16 (9)  &  7.65$\pm$0.13 (13) &  8.59$\pm$0.14 (9) \\
7862 &  2.0* & 8.48$\pm$0.12 (3)  &  7.86$\pm$0.08 (6)  &  8.95$\pm$0.10 (2) \\
7996 &  0.1* & 8.53$\pm$0.14 (3)  &  7.87$\pm$0.09 (7)  &  8.84$\pm$0.09 (4) \\
8385 &  1.2* & 8.57$\pm$0.19 (9)  &  7.97$\pm$0.12 (11) &  8.92$\pm$0.12 (8) \\
8549 &  1.6  & 8.26$\pm$0.07 (8)  &  7.74$\pm$0.07 (20) &  8.70$\pm$0.07 (23)\\
8768 &  1.4  & 8.42$\pm$0.10 (12) &  7.86$\pm$0.07 (20) &  8.88$\pm$0.16 (18)\\
9005 &  2.1  & 8.29$\pm$0.09 (10) &  7.84$\pm$0.08 (22) &  8.59$\pm$0.07 (21)\\
\hline
\end{tabular}
\end{minipage}
$^\ast$) These $V_t$ values are determined in Paper~III from He~I lines
\end{table}

As mentioned above, model atmospheres for the stars were computed using Kurucz's (1993) code ATLAS~9 for the derived parameters $T_{\rm eff}$ and $\log g$. When determining element abundances, apart from the star's model atmosphere the microturbulent parameter $V_t$ should be determined. We inferred the $V_t$ values from N~II and O~II lines. A standard procedure was used, namely: we searched for the $V_t$ that yields a zero slope for the abundance vs. $W$ relation; in other words, no trend of the N and O abundances with equivalent widths $W$ must exist. The derived parameters $V_t$ (N~II) and $V_t$ (O~II) were applied to the N and O abundance analyses, respectively. Since the few available  C~II lines  are weak, the C abundance is only very slightly dependent on $V_t$; we adopted the mean of the $V_t$~(N~II) and $V_t$~(O~II) -- see Table~5.
   For the  coolest programme stars, a reliable derivation of $V_t$ from N~II and O~II lines is impossible, because the few available lines are weak.  Therefore, for these stars (seven in all) we have taken $V_t$ from He~I lines (see Paper~III); these $V_t$ values are marked in Table~5 by asterisks.Table~5 shows that the $V_t$ values for the programme stars are rather small: they vary from 0.0 to 3.7 km~s$^{-1}$ with the average of 1.1 km~s$^{-1}$.  Of course, when all lines are weak, the derived abundance is independent of the adopted value of $V_t$.

      The derived C, N and O abundances with the standard deviations from the line-to-line scatter for 22 programme stars are presented in Table~5. The number of C~II, N~II and O~II lines used for each star is given there (in brackets). This number depends on the effective temperature $T_{\rm eff}$, and rotational velocity $v \sin i$, as well as the quality of the observed spectrum of a star. The C, N and O abundances are given in Table~5 as the mean values $\log\epsilon$ with the standard derivation $\sigma$. On average $\sigma$~=~$\pm$0.11 dex for C and $\pm$0.10 dex for N and O. Averaging data from Table~5, we obtain the following mean C, N and O abundances: $\log\epsilon$(C)~=~8.31$\pm$0.13, $\log\epsilon$(N)~=~7.80$\pm$0.12 and $\log\epsilon$(O)~=~8.73$\pm$0.13.

      As mentioned above, the accuracy of the C, N and O abundances depends on the reliability of the basic parameters $T_{\rm eff}$ and $\log g$. In order to estimate the accuracy, one may compare the derived mean $\log\epsilon$(C), $\log\epsilon$(N) and $\log\epsilon$(O) values with results obtained only for the stars with the most reliable parameters $T_{\rm eff}$ and $\log g$. First, there are 11 stars in Table~1 with the relatively small temperature errors $\Delta T_{\rm eff} < 500$~K; we found for them on average $\log\epsilon$(C)~=~8.34$\pm$0.13, $\log\epsilon$(N)~=~7.78$\pm$0.09 and $\log\epsilon$(O)~=~8.74$\pm$0.11. Second, the most accurate $\log g$ values are obtained for the stars with small errors in the parallaxes $\pi$. There are 7 programme stars with the minimum errors in $\pi$, i.e., errors of $\le$ 10 per cent. The mean abundances for these 7 stars are $\log\epsilon$(C)~=~8.28$\pm$0.10, $\log\epsilon$(N)~=~7.81$\pm$0.18 and $\log\epsilon$(O)~=~8.69$\pm$0.10. Third, there are 4 common stars from both previous groups, which have simultaneously the most accurate $T_{\rm eff}$ and $\log g$ values; we found for them $\log\epsilon$(C)~=~8.30$\pm$0.13, $\log\epsilon$(N)~=~7.74$\pm$0.08 and $\log\epsilon$(O)~=~8.72$\pm$0.13. One sees that in all three cases the mean C, N and O abundances virtually coincide with the above-mentioned averages for a total sample of 22 stars. It shoud be noted that van Leeuwen (2007) did not recommend use of parallaxes $\pi$ with errors $>$ 10 per cent. However, we found no marked difference in the mean C, N and O abundances between objects with errors in $\pi$ $>$ 10 per cent (15 stars) and $\le$ 10 per cent (7 stars): the difference is 0.04, 0.00 and 0.05 dex for C, N and O respectively, and these values are  less than uncertainties in the mean abundances.

      We estimated typical errors in the derived C, N and O abundances, which originate from four contributors, namely uncertainties in the parameters $T_{\rm eff}$, $\log g$ and $V_t$ and the scatter in the $\log\epsilon$ estimates from various lines. Since equivalent widths of C~II, N~II and O~II lines depend nonlinearly on $T_{\rm eff}$ (see, e.g., Figs 2 and 3), the contribution of the uncertainty in $T_{\rm eff}$ to a total error can be dependent on $T_{\rm eff}$. We selected a number of the stars from Table~1, which represent the whole $T_{\rm eff}$ range, and estimated for them the total errors in $\log\epsilon$(C), $\log\epsilon$(N) and $\log\epsilon$(O) adopting the following mean uncertainties: $\Delta T_{\rm eff}$~=~$\pm$500~K, $\Delta \log g$~=~$\pm$0.11 and $\Delta V_t$~=~$\pm$1.0 km~s$^{-1}$; the scatter in $\log\epsilon$ between lines is characterized by the mean dispersion $\sigma$~=~$\pm$0.10 dex (see above). We found that the total error in $\log\epsilon$(C) varies from 0.17 to 0.11 dex when $T_{\rm eff}$ varies from 15000 to 24000~K; the total error in $\log\epsilon$(N) and $\log\epsilon$(O) varies from 0.18 to 0.11 dex and from 0.21 to 0.13 dex, respectively. The contribution of the uncertainty $\Delta V_t$~=~$\pm$1.0 km~s$^{-1}$ seems to be imperceptible for all temperatures $T_{\rm eff}$ as the lines used are rather weak. The contribution of the $T_{\rm eff}$ uncertainty increases with decreasing $T_{\rm eff}$.  For the hottest stars of our sample ($T_{\rm eff}$ $\sim$ 22000-24000~K) the errors in $\log\epsilon$ originate mostly from the scatter in $\log\epsilon$ between lines, whereas for the cooler stars ($T_{\rm eff}$ $\sim$ 15000-19000~K) the contribution of the $T_{\rm eff}$ uncertainty is significant.

      Independent but limited confirmation of the accuracy of our C, N and O abundances is presented in Table~3, where our and Nieva \& Sim\'{o}n-D\'{i}az's (2011, hereinafter NS'11) data for two common stars from the Ori OB1 association are compared. The derived C, N and O abundances show very good agreement: differences in the $\log\epsilon$(C), $\log\epsilon$(N) and $\log\epsilon$(O) values vary from 0.00 to 0.06 dex, i.e. they are markedly less than errors of the $\log\epsilon$(C), $\log\epsilon$(N) and $\log\epsilon$(O) determination with only the C abundance for HR~1781 showing a greater difference, namely 0.12 dex.

      In the same vein, we  compare in Table 4 our C, N and O abundances with data of Nieva \& Przybilla (2012, hereinafter NP'12) for four common stars. 
In particular, one sees from Table 4 that the above-mentioned enhanced N abundance for HR~1810 is confirmed by NP'12. In general,
 an agreement between our and NP'12 results is worse than for two stars from NS'11 (Table 3). Especially large discrepancies with NP'12 are found for the C abundances in HR~2928 and HR~8385 (+0.26 and -0.24 dex, respectively). Note that in the case of HR~2928 the discrepancy is partially explained by the substantial difference between the adopted $T_{\rm eff}$ values.

      It should be noted that the C, N and O abundances derived in NS'11 and NP'12 show no correlation with the effective temperature $T_{\rm eff}$ (see Fig.7 in NP'12). Our N and O abundances show no dependence on $T_{\rm eff}$, too. However, unlike N and O, the C abundances for cooler stars of our sample with $T_{\rm eff}$~$\leq$~18000~K seem to be somewhat higher than for hotter stars. In fact, the mean value is $\log\epsilon$(C)~=~8.46$\pm$0.09 for the stars with $T_{\rm eff}$~$<$~18100~K (6 objects) and $\log\epsilon$(C)~=~8.25$\pm$0.08 for the stars with $T_{\rm eff}$~$>$~18500~K (16 objects). Note that there are only two relatively cool B stars in NS'11 and NP'12 with $T_{\rm eff}~<$~18000~K; their C abundances derived from C~II lines are in good agreement with the $\log\epsilon$(C) values for other stars. On the one hand, marked difference in the C abundances between the stars of our sample with $T_{\rm eff}~<$~18100~K and $T_{\rm eff}~>$~18500~K can result from some incompleteness of the used C model atom (in NS'11 and NP'12 the updated model is applied). On the other hand, it is interesting that the mean value $\log\epsilon$(C)~=~8.46 for the cooler stars agrees very well with the solar C abundance, whereas the mean value $\log\epsilon$(C)~=~8.25 for the hotter stars (as well as the mean C abundances in NS'11 and NP'12) show a marked underabundance.

\section{DISCUSSION}

\tabcolsep=0.09cm
\begin{table}
\caption{Comparison of our mean C, N and O abundances with other recent results}
\begin{minipage}{84mm}
\centering
\begin{tabular}{lcccc}
\hline
   Source                         &  Number    & $\log\epsilon$(C)   & $\log\epsilon$(N)   & $\log\epsilon$(O) \\
                                  & of stars   &                     &                     & \\
\hline                                                               
Present work                      &  22        & 8.31$\pm$0.13       & 7.80$\pm$0.12       & 8.73$\pm$0.13 \\
Takeda et al. (2010)              &  64        &       -             &       -             & 8.71$\pm$0.06 \\
Nieva \&                          &  13$^{a)}$ & 8.35$\pm$0.03       & 7.82$\pm$0.07       & 8.77$\pm$0.03 \\
 Sim\'{o}n-D\'{i}az (2011)        &            &                     &                     &\\
Cunha, Hubeny \&                  &  10$^{a)}$ &       -             &       -             & 8.78$\pm$0.05 \\
 Lanz (2012)                      &            &                     &                     &\\
Nieva \& Przybilla (2012)         &  29        & 8.33$\pm$0.04       & 7.79$\pm$0.04       & 8.76$\pm$0.05 \\
\hline
\end{tabular}
\end{minipage}
$^{a)}$ stars in the Ori OB1 association
\end{table}

      When our work was in progress, a several interesting publications appeared with C, N and O abundances in B stars. In Table 6, we compare our mean C, N and O abundances with these recent results. Some of the 64 B stars from Takeda et al.'s (2010) sample are close to or beyond the termination of the MS  (see their Fig.1). However, due to the relatively low masses ($M~=$~3-9~M$_\odot$) and slow rotational velocities ($v \sin i$ $\le$ 30 km~s$^{-1}$), these stars are expected to have retained their initial surface C, N and O abundances. Takeda et al. determined the O abundance from the O I 6156 \AA\ multiplet, and the Ne abundance from Ne~I lines but did not provide either C or N abundances. Of the  29 early B stars analysed by NP'12,  most are caught well before  the end of the  MS (see Fig.~5 there). An obvious exception is the star $\alpha$~Pyx (HR~3468, HD~74575) with the lowest surface gravity $\log g$~=~3.60, which is close to the MS's termination. Its mass is $M$ $\approx$ 12 M$_\odot$ and the projected rotational velocity is $v \sin i$~= 11~km~s$^{-1}$; according to  (Frischknecht et al. 2010) and assuming the star is not a rapid rotator observed at a high angle of inclination, such a star cannot change markedly its surface C, N and O abundances during the MS phase. On the whole, one may suppose that the vast majority of the C, N and O abundances presented in Table 6 correspond to the initial abundances in these young stars.

      Table 6 shows that our mean C, N and O abundances are in very good agreement with data from recent papers. In particular, there is excellent agreement with the mean C, N and O abundances found in NS'11 and NP'12. This is very important because both in the $T_{\rm eff}$ and $\log g$ determination and in the CNO abundance analysis quite different techniques were used by NS'11 (and NP'12) and us.

    It should be noted that NS'11, as well as Cunha, Hubeny \& Lanz (2012), studied the unevolved early B stars in the Ori OB1 association.  In both studies, the $T_{\rm eff}$ range of the stars is displaced to hotter temperatures relative to our sample: 19000-33400~K for NS'11 against 15360-24090~K for our sample. The agreement over abundances extends to Mg: 7.57$\pm$0.06 from NS'11 and 7.59$\pm$0.15 from Paper~IV. 

    Since the Ori OB 1 stars have been formed from interstellar material of the (presumed) same metallicity, it is quite natural that they show very homogeneous initial C, N and O abundances. In contrast, the scatter in the individual $\log\epsilon$(C), $\log\epsilon$(N) and $\log\epsilon$(O) values for the stars of our sample is greater than for the above samples from Ori OB 1. However, the stars of our sample do not belong to a single cluster or association; on the contrary, they are distributed around the solar neighbourhood up to $d$~=~600 pc. Therefore, we speculate that the spread among the individual $\log\epsilon$(C), $\log\epsilon$(N) and $\log\epsilon$(O) values for our stars  may be augmented by real star-to-star variation in the initial C, N and O abundances; the Galactic abundance gradient may be one source of the variation.

\tabcolsep=0.11cm
\begin{table}
\caption{The mean C, N and O abundances for the B-type MS stars in comparison with the solar abundances}
\begin{minipage}{84mm}
\centering
\begin{tabular}{lcccl}
\hline
Object      & $\log\epsilon$(C)       & $\log\epsilon$(N) & $\log\epsilon$(O) & Source \\
\hline
B-type stars&      8.31$\pm$0.13      &    7.80$\pm$0.12      & 8.73$\pm$0.13         & Present work \\
Solar       &      8.50$\pm$0.06      &    7.86$\pm$0.12      & 8.76$\pm$0.07         & Caffau et al. \\
photosphere &                         &                       &                       & (2008, 2009, 2010) \\
Solar       &      8.43$\pm$0.05      &    7.83$\pm$0.05      & 8.69$\pm$0.05         & Asplund et al. \\
photosphere &                         &                       &                       & (2009) \\
Protosolar  &      8.47$\pm$0.05      &    7.87$\pm$0.05      & 8.73$\pm$0.05         & Asplund et al. \\
abundances  &                         &                       &                       & (2009) \\
\hline
\end{tabular}
\end{minipage}
\end{table}

      In Table~7 our mean $\log\epsilon$(C), $\log\epsilon$(N) and $\log\epsilon$(O) values for 22 B stars are compared with the recent  solar values  (Caffau et al. 2008, 2009, 2010;  Asplund et al. 2009). Protosolar abundances are 0.04 dex larger (Asplund et al. 2009). When comparing the results in Table~7, we conclude that our mean $\log\epsilon$(N) and $\log\epsilon$(O) values for the B stars coincide within errors of their determination with the photospheric and protosolar abundances. However, our mean $\log\epsilon$(C) value, also those in NS'11 and NP'12 (Table 6), shows a marked underabundance in comparison with the Sun (the deficiency is especially noticeable when compared with small errors in the mean C abundances in NS'11 and NP'12).

\begin{figure}
\begin{center}
\includegraphics[width=84mm]{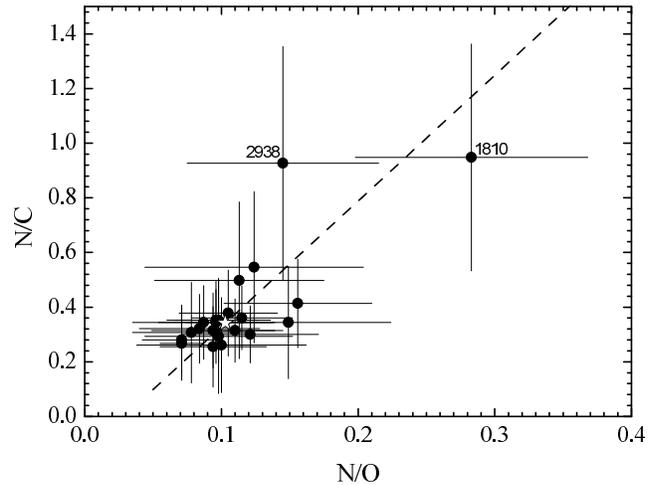}
\end{center}
\caption[]{Relation between the mass ratios N/C and N/O for 22 programme stars. Two stars with the high N/C values, namely HR~1810 and HR~2938, are marked. Theoretical path of the N/C vs. N/O relation is shown by the dashed line drawn through the initial values N/O~=~0.103 and N/C~=~0.342 (open star in the figure) with a slope d(N/O)/d(N/C)~=~4.6 from NP'12.}
\end{figure}
      
      We also determined the ratios N/O and N/C for 22 programme stars; we found the mean values $\log$~(N/O)~=~-0.91$\pm$0.13 and $\log$~(N/C)~=~-0.50$\pm$0.15. The first value is very close to the solar ratio $\log$~(N/O)~=~-0.90$\pm$0.14 (Caffau et al.) or -0.86$\pm$0.07 (Asplund et al.). The second one is somewhat greater than the solar ratio $\log$~(N/C)~=~-0.64$\pm$0.13 (Caffau et al.) or -0.60$\pm$0.07 (Asplund et al.); a main cause of this difference is a slight C deficiency in B stars.

      Following to NP'12 (see their Fig.14), we show in Fig.9 the N/C vs. N/O diagram, where the derived CNO abundances, like NP'12, have been converted into the mass fraction scale. One may see that there is no correlation between N/O and N/C for the most of the stars; this is explainable because their observed CNO abundances are expected to correspond to the unchanged initial values. However, two of the stars, namely HR~1810 and HR~2938, show the relatively high N/C values. One may suppose that these two stars could change their atmospheric abundances of light elements during the MS phase evolution due to the rotationally induced mixing. However, their positions in the $T_{\rm eff}$ -- $\log g$ diagram (Fig.8) are not exclusive. Moreover, the presented errors in the N/O and N/C values for these stars are large (note that the errors in Fig.9 are proportional to N/O and N/C themselves), so their difference from other stars cannot recognize as a reliable one. Nevertheless, it should be noted that the relatively high N/C ratios for HR~1810 and HR~2938 were confirmed by NP'12.

      We also display in Fig.9 a possible theoretical trend (dashed line) which corresponds to the NP'12 slope d(N/O)/d(N/C)~=~4.6 and our initial values N/O~=~0.103 and N/C~=~0.342 found for 20 stars without HR~1810 and HR~2938 (open star in Fig.9); note that these initial values are very close to the NP'12 ones. When excluding for the assumptive mixed stars HR~1810 and HR~2938, we obtain the following mean abundances for the remaining 20 stars: $\log\epsilon$(C)~=~8.33$\pm$0.11, $\log\epsilon$(N)~=~7.78$\pm$0.09 and $\log\epsilon$(O)~=~8.72$\pm$0.12. We compare in Table 8 these values with a present-day Cosmic Abundance Standard (CAS) of NP'12; one sees that there is an excellent agreement.

\begin{table}
\caption{Comparison of our mean C, N and O abundances for 20 B stars (the 'mixed' stars HR~1810 and HR~2928 are          omitted) with a present-day Cosmic Abundance Standard (CAS) from NP'12}
\centering
\begin{tabular}{ccc}

  \hline
  Element & Our abundances & CAS \\
          &                &(NP'12)\\
  \hline
  C & 8.33$\pm$0.11 & 8.33$\pm$0.04 \\
  N & 7.78$\pm$0.09 & 7.79$\pm$0.04 \\
  O & 8.72$\pm$0.12 & 8.76$\pm$0.05 \\
  \hline

\end{tabular}
\end{table}

      Thus, our findings, as well as up-to-date data of other authors show that the observed (unevolved) N and O abundances in B stars are very close to the solar ones. On the other hand, carbon shows a small stable deficiency. In fact, using the old carbon model atom of Sigut (1996), we found the mean abundance $\log\epsilon$(C)~=~8.31$\pm$0.13. Basing on the updated carbon model atom, NS'11 and NP'12 obtained the very close values 8.35$\pm$0.03 and 8.33$\pm$0.04, respectively (note that errors are very small there). All these values are markedly lower than the current $\log\epsilon$(C) evaluations for the Sun (Table 7). It is important to note that many other chemical elements show abundances that are in an excellent agreement with the solar ones (see Introduction). Why does only carbon show such a unique difference from the Sun's abundance?

      It is possible that there are some specific details in non-LTE computations of C~II lines that were ignored up to now. In particular, the C~II ionization energy is $E_{\rm ion}$~=~24.383 eV that corresponds to the wavelength $\lambda$~=~508.5 \AA; therefore, the C~II photoionization is controlled by the EUV radiation. For example, such a radiation was observed for the star $\epsilon$~CMa (B2 II); it was found that the observed EUV flux at $\lambda~<$~504~\AA\ (He~I continuum) is significantly greater than the computed one (Cassinelli et al. 1995). Later Gregorio at al. (2002), using both the ATLAS9 model atmospheres of Kurucz (1993) and the spherical model atmospheres of Aufdenberg et al. (1999), showed that a more satisfactory agreement with the observed EUV flux for the star $\epsilon$~CMa can be achieved; however, the observed flux has a different shape than the predicted one. Moreover, the observed line intensities in this region seem to be much stronger than predicted ones. Thus, the theory cannot describe adequately the EUV flux which is important for computations of the C~II photoionization (as well as the N~II and O~II photoionization, too). Moreover, we cannot exclude that some further update of the carbon model atom is needed to solve the problem.

      Another alternative was proposed recently; it was supposed that the Sun can migrate during its life (i.e., for the past $4.5\cdot 10^9$~yr) from inner parts of the Galactic disk where it has born, so its observed chemical composition may differ from the composition of young stars in its present neighbourhood (typical ages of the B stars in question are $t \sim (10-100)\cdot 10^6$~yr). This hypothesis is discussed, e.g., in NP'12 (see also references there). Detailed analysis of the radial migration of stars (including the Sun) within the Galaxy is very hard; nevertheless, the question remains: why is it that only the carbon abundance shows a marked difference from the Sun's abundances? 
In our opinion, the slight carbon deficiency in the unevolved B stars is a subject for further discussion.

\section{CONCLUDING REMARKS}

We selected from our original list of early and medium B-type stars (Paper~III) the relatively cool stars with low rotational velocities ($v \sin i < 70$~km~s$^{-1}$). Their two basic parameters, namely the effective temperature $T_{\rm eff}$ and the surface gravity $\log g$ were redetermined; application of new stellar parallaxes allowed increasing markedly the accuracy of the derived $\log g$ values. Our final list contains 22 stars with the following parameters: (i) effective temperatures $T_{\rm eff}$ are less than 24100 K; (ii) surface gravities $\log g$ are mostly greater than 3.75; (iii) the projected rotational velocities are $v \sin i \le 66$~km~s$^{-1}$. These stars with moderate masses $M$~=~5-11 M$_\odot$ and slow rotation should keep their initial surface C, N and O abundances according to the theory of rotationally-induced mixing.

      We found for these stars the mean abundances $\log\epsilon$(C)~=~8.31$\pm$0.13, $\log\epsilon$(N)~=~7.80$\pm$0.12 and $\log\epsilon$(O)~=~8.73$\pm$0.13; we suppose that these values are the initial CNO abundances in B-type MS stars. The derived N and O abundances coincide with the solar values, whereas the C abundance is somewhat lower than the solar one. We showed that these C, N and O abundances are in good agreement with the recent values for B-type MS stars by other authors. In particular, the small C deficiency is confirmed by the recent studies by  NS'11 and NP'12.

      Two programme stars, namely HR~1810 and HR~2938, can be mixed during the MS phase. If these stars are omitted, we obtain the following mean abundances for the remaining 20 stars: $\log\epsilon$(C)~=~8.33$\pm$0.11, $\log\epsilon$(N)~=~7.78$\pm$0.09 and $\log\epsilon$(O)~=~8.72$\pm$0.12; these values are in excellent agreement with a present-day Cosmic Abundance Standard (CAS) of Nieva \& Przybilla (2012).

      Thus, on the one hand, the initial N and O abundances in nearby B-type stars, as well as the abundances of some other metals, confirm that young stars in the solar neighbourhood and the Sun have the same metallicity. In the absence of a migration of the Sun from its birthplace, one may conclude that during the Sun's life (i.e., for the past $4.5\cdot10^9$~yr) the metallicity of the solar neighbourhood has not markedly changed; so, an intensive enrichment of the solar neighbourhood by metals occurred before the Sun's birth. On the other hand, carbon is an evident 'outlier' in this scenario, because the observed C abundance in B stars shows  a small deficiency in comparison with the Sun. This carbon deficiency demands an alternative hypothesis  either the Sun has migrated from inner parts of the Galactic disk where it has born or present non-LTE computations overlook one or more physical effects in the carbon ions. The C underabundance in B stars needs further study.

\section*{Acknowledgments}

We thank Sergey Korotin for the updated code MULTI and input data for non-LTE computations of C~II, N~II and O~II lines. We thank Maria Nieva for interesting discussions. DLL acknowledges the support of the Robert A. Welch Foundation of Houston, Texas through grant F-634.

\label{lastpage}
\end{document}